\documentclass[10pt]{article}

\usepackage{amsmath}
\usepackage{amssymb}

\usepackage{graphicx}

\usepackage{cite}
\usepackage{color} 

\usepackage[labelfont=bf,labelsep=period,justification=raggedright]{caption}

\makeatletter
\renewcommand{\@biblabel}[1]{\quad#1.}
\makeatother
\newcommand{\rem}[1]{}

\date{}

\begin{document}

\begin{flushleft}
{\Large
\textbf{Bistable forespore engulfment in \textit{Bacillus subtilis} by a zipper mechanism in absence of the cell wall}
}
\\
Nikola Ojkic$^{1, 2, \ast}$, 
Javier L\'{o}pez-Garrido$^{3}$,
Kit Pogliano$^{3}$,
Robert G. Endres$^{1, 2}$, 

\bf{1} Department of Life Sciences, Imperial College, London, United Kingdom
\\
\bf{2} Centre for Integrative Systems Biology and Bioinformatics, Imperial College, London, United Kingdom
\\
\bf{3} Division of Biological Sciences, University of California at San Diego, La Jolla, California, USA
\\
$\ast$ E-mail: n.ojkic@imperial.ac.uk
\end{flushleft}

\date{\today}

\section*{Abstract}

To survive starvation, the bacterium \textit{Bacillus subtilis} forms durable spores. The initial step of sporulation is asymmetric cell division, leading to a large mother-cell and a small forespore compartment. After division is completed and the dividing septum is thinned, the mother cell engulfs the forespore in a slow process based on cell-wall degradation and synthesis. However, recently a new cell-wall independent mechanism was shown to significantly contribute, which can even lead to fast engulfment in $\sim$ 60 $\%$ of the cases when the cell wall is completely removed. In this backup mechanism, strong ligand-receptor binding between mother-cell protein SpoIIIAH and forespore-protein SpoIIQ leads to zipper-like engulfment, but quantitative understanding is missing. In our work, we combined fluorescence image analysis and stochastic Langevin simulations of the fluctuating membrane to investigate the origin of fast bistable engulfment in absence of the cell wall. Our cell morphologies compare favorably with experimental time-lapse microscopy, with engulfment sensitive to the number of SpoIIQ-SpoIIIAH bonds in a threshold-like manner. By systematic exploration of model parameters, we predict regions of osmotic pressure and membrane-surface tension that produce successful engulfment. Indeed,  decreasing the medium osmolarity in experiments prevents engulfment in line with our predictions. Forespore engulfment may thus not only be an ideal model system to study decision-making in single cells, but its biophysical principles are likely applicable to engulfment in other cell types, e.g. during phagocytosis in eukaryotes.

\section*{Author Summary}

When the bacterium \textit{B. subtilis} runs out of food, it undergoes a fundamental development process by which it forms durable spores. Sporulation is initiated by asymmetric cell division after which the larger mother cell engulfs the smaller forespore, followed by spore maturation and release. This survival strategy is so robust that engulfment even proceeds when cells are deprived of their protective cell wall. Under these severe perturbations, 60 $\%$ of the mother cells still engulf their forespores in only 10 $\%$ of the normal engulfment time, while the remaining 40 $\%$ of mother cells withdraw from engulfment. This all-or-none outcome of engulfment suggests decision-making, which was recently also identified in other types of engulfment, e.g. during phagocytosis when immune cells engulf and destroy pathogens. Here, we developed a biophysical model to explain fast bistable forespore engulfment in absence of the cell wall and energy sources. Our discovered principles may prove very general, thus predicting key ingredients of successful engulfment across all kingdoms of life.

\section*{Introduction}

\begin{figure}[!ht]
\begin{center}
\includegraphics[width=15cm, trim = 0cm 0cm 0cm 0cm, clip=true]{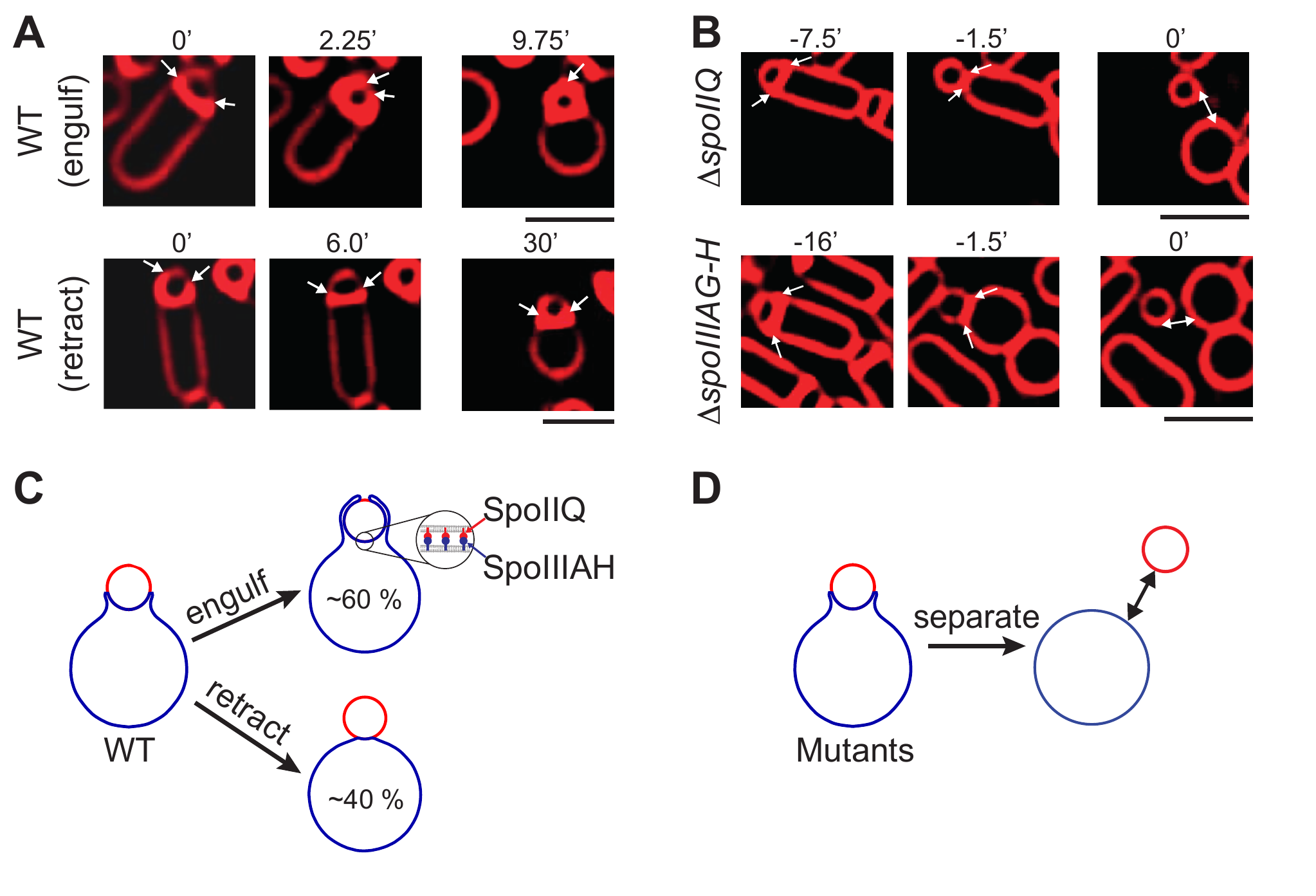}
\end{center}
\caption{
{\bf Bistable forespore engulfment of \textit{B. subtilis} after cell-wall removal.} (A and B) Images adopted from \cite{Broder:2006}. Medial focal plane images of sporulating bacteria treated with cell-wall removal lysozyme in osmotically protected medium with 0.5 M of sucrose \cite{Broder:2006}. Fluorescent membrane stain FM 4-64 was used to track the progressing mother-cell membrane engulfing the forespore. Arrows point to the moving edges of the mother membrane. Double-headed arrows show mother-forespore cell separation. (A) In wild-type (WT) cells upon cell-wall removal, mother cell  either engulfs the forespore (top) or retracts (bottom), see Movie S1. Time 0 minutes (0') is assigned to the onset of volume loss (see Fig. \ref{Fig_02}A). (B) Absence of the zipper proteins SpoIIQ (top) or SpoIIIAH (bottom) prevents membrane from forward progression causing protoplast separation. Time 0 minutes is used as the time of physical separation of mother cell and forespore. (C) Cartoon of fast bistable forespore engulfment in WT cells. Mother-cell compartment and forespore are shown in blue and red, respectively. After cell-wall removal $\sim$ 60 $\%$ of the sporulating cells engulf the forespore, while $\sim$ 40 $\%$ fail to engulf \cite{Broder:2006}. (D) Cartoon showing the protoplast separation as observed in mutants of panel B. Scale bars: 2 $\mu$m.
}
\label{Fig_01}
\end{figure}

To survive starvation the Gram-positive bacterium \textit{Bacillus subtilis} develops durable spores among other survival strategies \cite{Tan2014}. During sporulation, bacteria go through a costly developmental process under limited energy resources. The initial morphological step of sporulation is asymmetric cell division, resulting in a large mother-cell and a small forespore compartment \cite{Stragier:1996}. Subsequently, the  dividing septum is largely degraded and the mother-cell membrane moves around the forespore. This membrane movement is similar to phagocytosis whereby immune cells clear our bodies from pathogens and other particles \cite{Swanson:2008,Underhill:2002}. Finally, the engulfed forespore matures into a spore and the mother cell lyzes for its release. The origin of the engulfment force has been a topic of current research \cite{Abanes:2002,Blaylock:2004,Doan:2005,Broder:2006,Morlot:2010,Meyer:2010,Rodrigues:2013}. Cell-wall degradation and new cell-wall deposition were shown to play a significant role in advancing the mother-cell membrane leading edge. Strikingly, when the cell wall is enzymatically removed engulfment still occurs, surprisingly taking only $\sim$ 1-2 min compared to $\sim$ 45 min with the cell wall (see Fig. \ref{Fig_01}, Movie S1)\cite{Broder:2006}. Furthermore, engulfment is successful in $\sim$ 60 $\%$ of cells while the remaining $\sim$ 40 $\%$ retract. This observation raises questions on the origin of bistability and decision-making in relatively simple systems under severe energy limitations.

In the absence of the cell wall, migration of the mother-cell membrane around the forespore depends on the two membrane proteins that bind each other with high affinity \cite{Meisner:2011}, constituting a backup mechanism under severe perturbations \cite{Broder:2006,Gutierrez2010}: SpoIIQ expressed in the forespore and SpoIIIAH expressed in the mother cell \cite{Broder:2006,Levdikov:2012,Meisner:2012}(see Fig. \ref{Fig_01}A-D). To facilitate engulfment a physical mechanism similar to a Brownian ratchet was proposed \cite{Broder:2006}. Specifically, thermal fluctuations move the leading membrane edge forward, thus establishing new SpoIIQ-SpoIIIAH bonds that prevent backward membrane movement. One striking feature, however, is that the membrane cup surrounding the forespore is very thin (Fig. \ref{Fig_01}A, top). This either indicates a fast nonequilibrium mechanism for engulfment or additional forces that produce high membrane curvatures around the cup's neck region. Even though modeling of similar processes such as membrane budding and phagocytosis helped us understand the role of physical constraints on engulfment \cite{Herant:2006,Van:2009,Tollis:2010,Herant:2011}, quantitative modeling of forespore engulfment as a fundamental development process is still missing.

Here, using image analysis, Langevin simulations and simple analytical approaches we show that fast forespore engulfment in the absence of the cell wall occurs below $\sim$ 1 min, consistent with out-of-equilibrium dynamics driven by strong SpoIIQ-SpoIIIAH binding. Furthermore, we find physical parameter regimes responsible for bistable engulfment, including the number of bonds necessary for threshold-like engulfment and suitable osmotic pressures. The former prediction matches previously published data, while we successfully tested the latter with time-lapse microscopy. Hence, our model makes testable predictions on the measurable physical parameters leading to fast, energy-efficient engulfment. Forespore engulfment in the absence of the cell wall is thus  an ideal system to study phagocytosis-like processes and decision-making in single-cell organisms.

\section*{Results}
\subsection*{Image analysis reveals drastic mother-cell volume loss}
To better understand the process of engulfment in the absence of the cell wall, we analyzed the volume and surface area of sporulating cells treated with cell-wall removal enzyme (lysozyme) from previously published data \cite{Broder:2006}. For this purpose we used the semi-automated image-analysis software JFilament \cite{Smith:2010} (see Materials and Methods). Briefly, JFilament software allows assisted manipulation of active contours that comply with the bright linear structures of the images. Therefore, membranes that were fluorescently stained with FM 4-64 were easily tracked over time. Information about cell-membrane position in the medial focal plane was used to calculate volume and surface area assuming rotational symmetry around the axis connecting mother and forespore center of mass.

\begin{figure}[!ht]
\begin{center}
\includegraphics[width=15cm, trim = 0cm 0cm 0cm 0cm, clip=true]{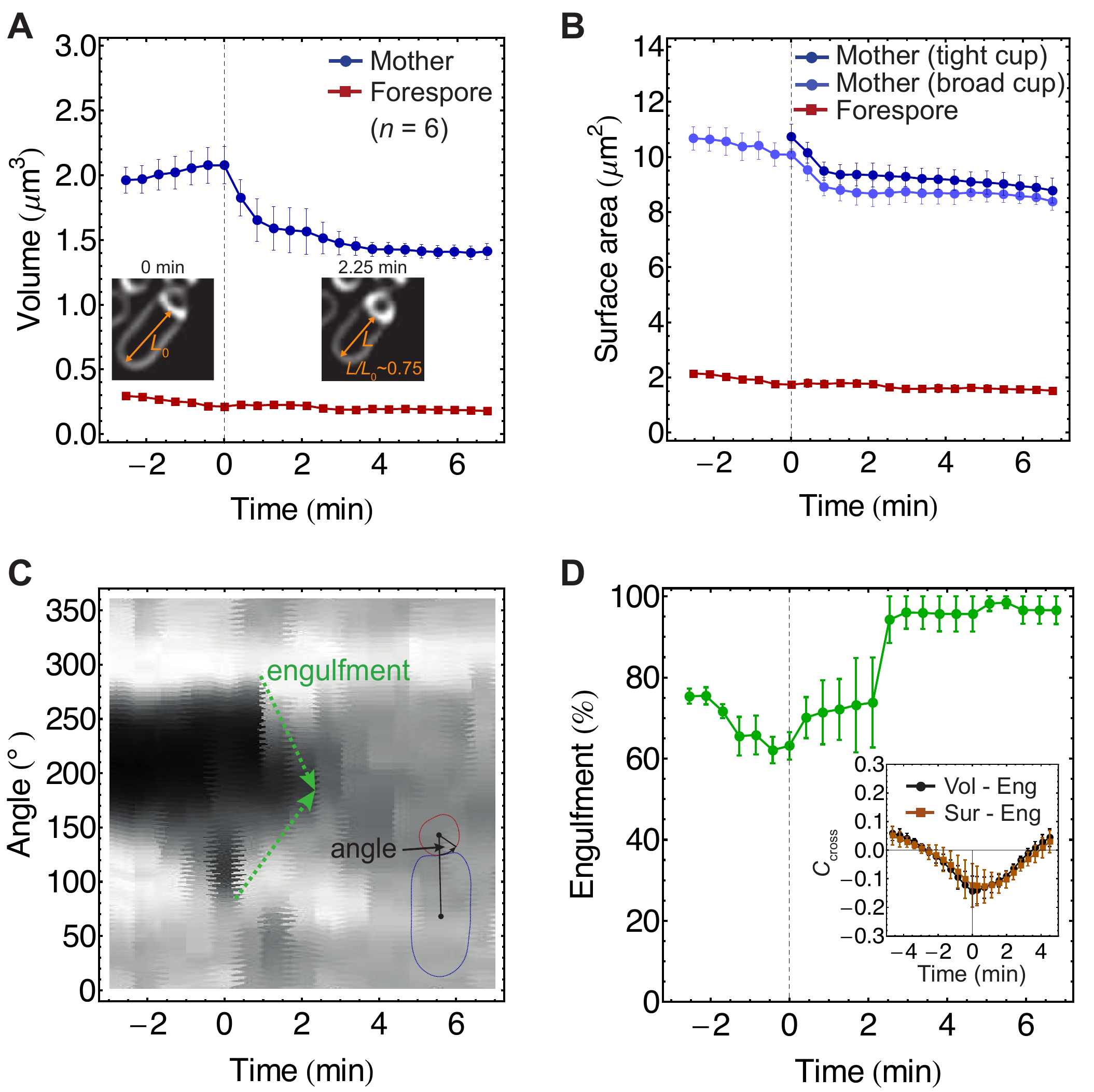}
\end{center}
\caption{
{\bf Image analysis reveals mother-cell volume loss during engulfment.} (A-D) Using active contours  we measured volume, surface area, and engulfment over time for mother cells and forespores (for details see Materials and Methods). The onset of volume loss for each cell was set to 0 minutes and all cell measurements were aligned in time based on 0' points. 3D volume and surface area were calculated assuming rotational symmetry around the axis that connects center of masses of forespore and mother cells. All analysis was performed on previously published movies from \cite{Broder:2006}. (A) Mother-cell volume loss amounts to $\sim$ 35 $\%$ during engulfment, while forespore volume remains the same. (Inset) Typical cell shrinks longitudinally causing volume loss. (B) Surface area for mother cell was calculated using two models, termed ``tight cup'' or ``broad cup'' shown in Fig. \ref{Fig_03}A. The broad-cup model is assumed for times $<$ 0' as no morphological membrane changes occur for these times (see Fig. \ref{Fig_03}B). The mother-surface area reduction is $\sim$ 5-15 $\%$, while forespore-surface area remains the same. (C) Engulfment shown by FM 4-64 kymograph (pixel intensities along the forespore contour versus time) of a single representative cell treated with lysozyme coincides with onset of volume loss. (D) Average engulfment over time. (Inset) Cross-correlation coefficient versus time showing anticorrelation between mother-cell volume and engulfment, and mother-surface area and engulfment. No time delays were observed. Average +/- SEM of $n$ = 6 cells. 
}
\label{Fig_02}
\end{figure}

Upon cell-wall removal, a drastic volume loss of $\sim$ 35 $\%$ in engulfing mother cells was observed (see Fig 2A). The onset of volume loss for each cell was set to 0 min, and surface area and engulfment measurements were aligned in time based on this time point. Similar analyses showed that no changes in forespore volume were observed (Fig. \ref{Fig_02}A). However, surface-area analysis during engulfment did reveal a minor reduction of the mother-cell surface by $\sim$ 5-15 $\%$ (Fig. \ref{Fig_02}B), depending on the assumed shapes of the progressing membrane in the image analysis. To correlate the drastic volume loss with the onset of engulfment we created kymographs along the fluorescently labeled forespore membranes of the engulfing cells (Fig. \ref{Fig_02}C and D). Extracting from the kymographs the percentage of forespore engulfment over time revealed that engulfment in the absence of the cell wall occurs on a time scale of $\sim$ 1-2 min as opposed to $\sim$ 45 min with the cell wall \cite{Broder:2006,Meyer:2010}.

Calculating cross-correlations we found significant anticorrelations between volume and engulfment, and surface area and engulfment (Fig. \ref{Fig_02}D). No time delays were detected (the minimum of the cross-correlation function is at 0 min) at a sampling rate of 1.3 frame/min (for details see Materials and Methods). However, under the same experimental conditions no volume or surface-area losses were observed in cells lacking SpoIIQ or SpoIIIAH zipper-molecules after cell-wall removal (see Fig. S1). Therefore, this led us to conclude that the drastic volume loss may play a facilitating role in engulfment.

\subsection*{Cup shape dynamically changes from broad to thin}
\begin{figure}[!ht]
\begin{center}
\includegraphics[width=15cm, trim = 0cm 0cm 0cm 0cm, clip=true]{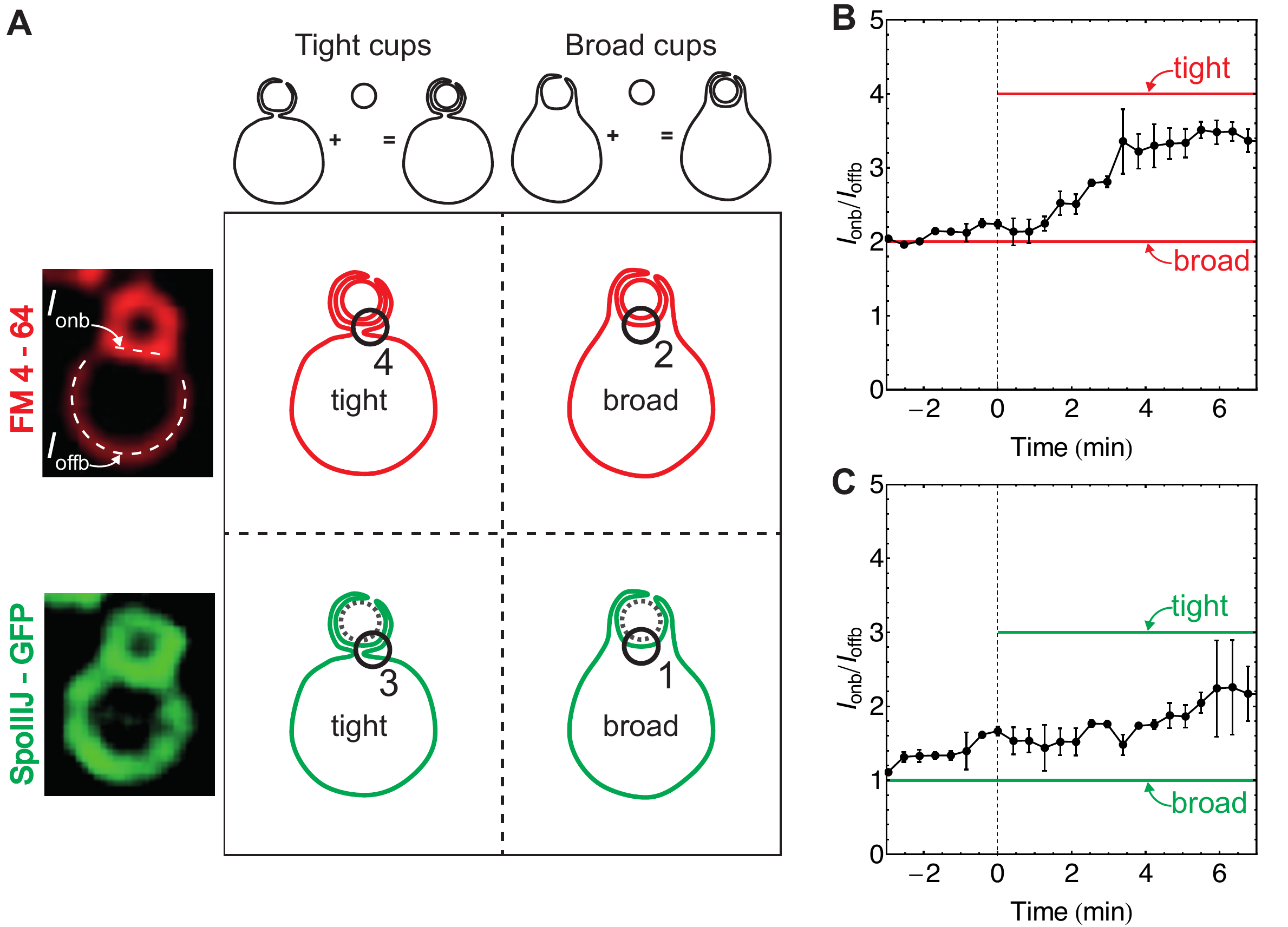}
\end{center}
\caption{
{\bf Cup-shape analysis.} (A) Theoretical predictions of two exclusive cup-shape models in two different fluorescence channels. Fluorescent FM 4-64 labels all membrane exposed to medium, while SpoIIIJ-GFP localizes at the mother-cell membrane only \cite{Broder:2006}. Therefore, the ``tight-cup'' model predicts four membrane folds at the mother-forespore boundary in the FM 4-64 channel and three membrane folds observed in the GFP channel. Likewise, ``broad-cup'' model predicts two membrane folds in the FM 4-64 channel and a single membrane in the GFP channel. (B-C) For a single cell at certain time points two average intensities were measured: average pixel intensity of mother-forespore boundary ($I_{\rm onb}$), and average pixel intensity of mother membrane far from the boundary ($I_{\rm offb}$). Ratio $I_{\rm onb}/I_{\rm offb}$ versus time is plotted for FM 4-64 channel (B) and for SpoIIIJ-GFP (C) . As before, time 0' is the onset of volume loss (see Fig. \ref{Fig_02}). Average +/- SEM for different cells is plotted. See Movie S2.}
\label{Fig_03} 
\end{figure}

During fast engulfment the mother-cell membrane drastically changes shape. Based on fluorescence intensity measurements, a very thin and tight cup was proposed at the late stage of engulfment  $\sim$ 20 min after cell-wall removal (Fig. \ref{Fig_03}A) \cite{Broder:2006}. However, this tight cup amounts to very high membrane curvatures. Therefore, we wondered how the thin cup with high membrane-bending energy may emerge from an initial broad cup, as observed in cells prior to cell-wall removal (Fig. \ref{Fig_01}A, top). To answer this question, we analyzed fluorescence images in two channels: fluorescent FM 4-64 (red) that uniformly binds  to all membranes exposed to the medium, and SpoIIIJ-GFP (green), a labeled protein that is only expressed in the mother cell where it is uniformly recruited to the cell membrane \cite{Broder:2006}.

In Fig. \ref{Fig_03}A the tight-cup model predicts four membrane folds at the mother-forespore boundary in the FM 4-64 channel and three membrane folds in the GFP channel. Likewise, the broad cup model predicts two membrane folds in the FM 4-64 channel and a single membrane in the GFP channel. To quantify the transition from a broad to the proposed thin cup we measured the average intensity of the mother-forespore boundary ($I_{\rm onb}$) and mother-cell membrane intensity far from the boundary ($I_{\rm offb}$, see Fig. \ref{Fig_03}A). Ratio $I_{\rm onb}/I_{\rm offb}$ is then used to determine the number of membrane folds on the mother-forespore interface. Relative boundary analysis of FM 4-64 intensity revealed that during late-stage engulfment cup shape conforms in between a broad and tight cup with $\sim$ 3.5 membrane folds on the mother-forespore interface. Similarly, analysis of the noisy GFP channel (Fig. \ref{Fig_03}C, and Movie S2) suggests that a broad cup indeed undergoes a transition towards a thin cup through significant morphological changes at the onset of engulfment. To quantitatively explain the observed volume loss, membrane morphological changes, and fast bistable engulfment, we implemented a biophysical model of forespore engulfment using Langevin dynamics.

\subsection*{Biophysical model of engulfment}
We hypothesized that thin cups form due to fast nonadiabatic engulfment away from equilibrium. If, however, it turns out that the engulfment dynamics are not fast enough under biophysical membrane constraints, then additional forces would need to be postulated to produce such high-curvature membrane features. To test our hypothesis we used Langevin dynamics that account for out-of equilibrium processes. 

\begin{figure}[!ht]
\begin{center}
\includegraphics[width=12cm, trim = 0cm 0cm 0cm 0cm, clip=true]{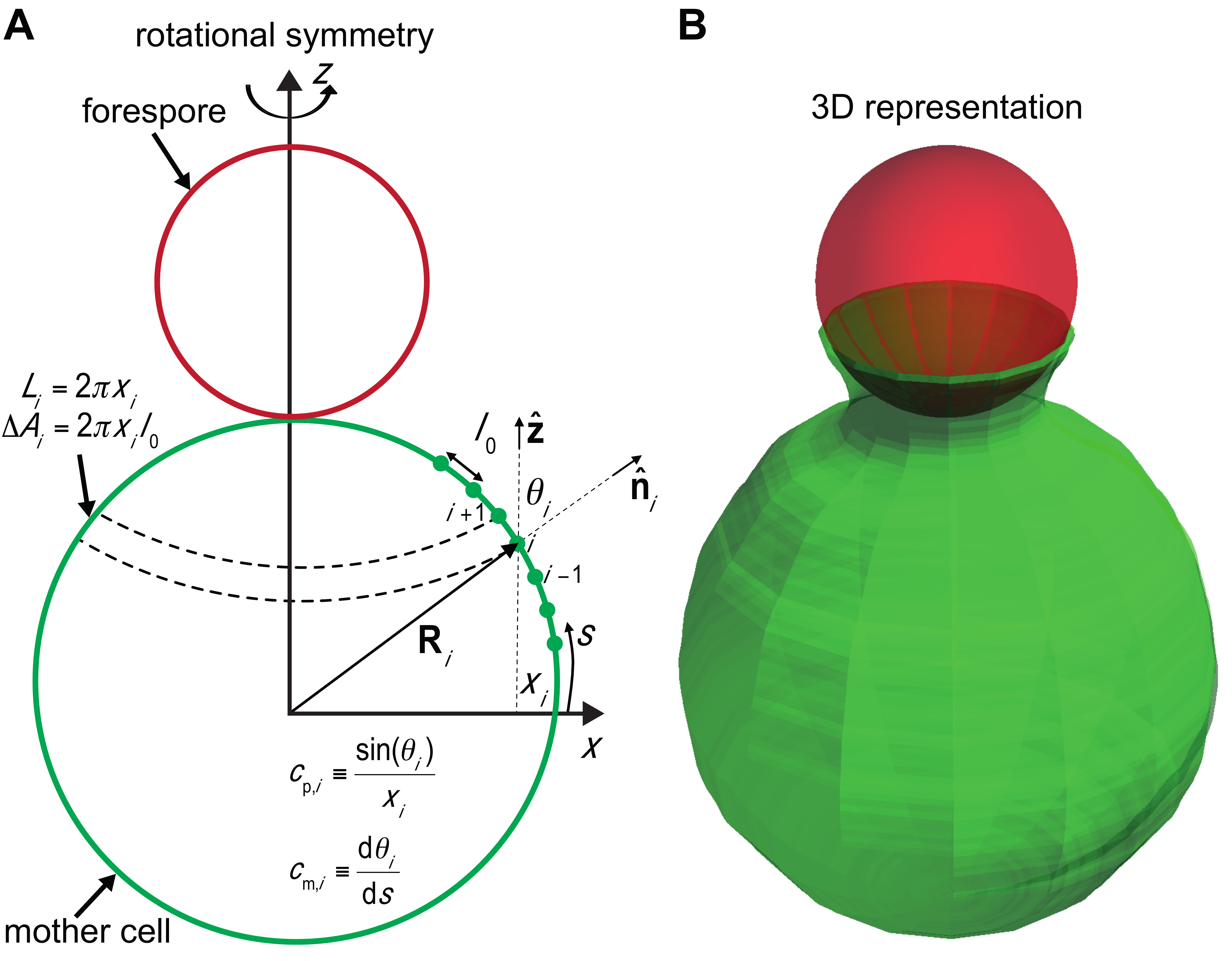}
\end{center}
\caption{
{\bf Engulfment model and Langevin simulations.} (A) 3D mother-cell membrane is represented by a string of beads assuming rotational symmetry around $z$-axis. Each bead at position ($x_{i}$, $z_{i}$) represents a ribbon of width $l_{0}$ and length $L_{i}$. Forespore is modeled as a solid sphere. See Materials and Methods, and Text S1 for further model explanations. (B) Snapshot of 3D simulation showing example of early-stage engulfment.
}
\label{Fig_04}
\end{figure}

In our model the 3D mother-cell membrane is represented by a string of beads assuming rotational symmetry around the $z$-axis, while the forespore membrane was modeled as a hard sphere (Fig. \ref{Fig_04}). Indeed, experiments show negligible deformation of the forespore during engulfment (Fig. \ref{Fig_02}A and B). Specifically, the Langevin dynamic equation of the $i^{\rm th}$ bead at  position $\textbf{R}_i$ is given by: 
\begin{equation}
\zeta_{i}\frac{\mathrm d \textbf{R}_i}{\mathrm d t}=\textbf{F}^{\mathrm {bend}}_{i}+\textbf{F}^{\mathrm {stoch}}_{i}+\textbf{F}^{\mathrm {QAH}}_{i}+\textbf{F}^{\sigma}_{i}+\textbf{F}^{\Delta p}_{i},
\label{Langevin1}
\end{equation}
where each bead at the position $\textbf{R}_i$ = ($x_{i}$, $z_{i}$) represents a ribbon of width $l_{0}$ and length $L_{i}$ shown in Fig. \ref{Fig_04}A. The left-hand side of Eq. 1 depends on the drag coefficient $\zeta_{i} \approx 4 \pi \eta_{\rm med} L_{i}$ \cite{Howard:2001}, with $\eta_{\rm med}$ is the effective medium viscosity (see Text S1). On the right-hand side of Eq. \ref{Langevin1} we have contributions of membrane bending, stochastic thermal fluctuations, zipper-molecule binding, surface tension, and osmotic pressure. For each term we give a brief description, while the detailed model equations and analytical derivations can be found in  Materials and Methods and Text S1, respectively.

The membrane-bending force ($\textbf{F}^{\mathrm {bend}}_{i}$) restores the curved membrane to the equilibrium flat configuration. The stochastic term ($\textbf{F}^{\mathrm {stoch}}_{i}$) is used to simulate thermal fluctuations with an effective temperature representing a driving force for the leading-edge ratchet movement. The amplitude of the membrane thermal fluctuations is chosen to be $\approx$ 15 nm (see Fig. S2) as typically observed for lipid bilayers \cite{Sackmann:1996} and red-blood cells \cite{Betz:2009}. The zipper-molecule force ($\textbf{F}^{\mathrm {QAH}}_{i}$) accounts for the high binding affinity between SpoIIQ and SpoIIIAH \cite{Meisner:2012}. This zipper mechanism is a strong driving force of the membrane leading edge. For the zipper-protein surface density ($\rho$) we initially chose the maximum possible value that corresponds to a single molecule per 100 nm$^2$ of membrane (based on $\sim$ 10 nm for the size of the protein \cite{Meisner:2012,Levdikov:2012} and assumed dense packing). The surface-tension force ($\textbf{F}^{\sigma}_{i}$) is characterized by linear ($\sigma_1$) and nonlinear ($\sigma_2$) surface tensions (see Materials and Methods) \cite{Herant:2006}. The force due to osmotic pressure ($\textbf{F}^{\Delta p}_{i}$) is characterized by the pressure difference ($\Delta p$) between inner and outer medium. While the surface-tension term causes membrane contraction, the pressure-difference term produces volume expansion. In thermal equilibrium these two forces balance each other. An example of an implemented simulation is shown in Fig. \ref{Fig_04}B. 

To validate our simulations we considered a spherical mother cell prior to engulfment, allowing us to quantify membrane fluctuations in thermal equilibrium. Specifically, we used the Langevin equation without the zipper term $\textbf{F}^{\mathrm {QAH}}_{i}$ on the right-hand side of Eq. \ref{Langevin1}. Obtained numerical results were compared with analytical results of thermal membrane fluctuations \cite{Meleard:1992,Milner:1987} and fluctuation spectra \cite{Pecreaux:2004} (Fig. S3). This validation showed that our 3D model of the mother cell has indeed appropriate membrane biophysical properties. For further details see Materials and Methods.

\subsection*{Simulation of fast engulfment produce thin cups}

\begin{figure}[!ht]
\begin{center}
\includegraphics[width=15cm, trim = 0cm 0cm 0cm 0cm, clip=true]{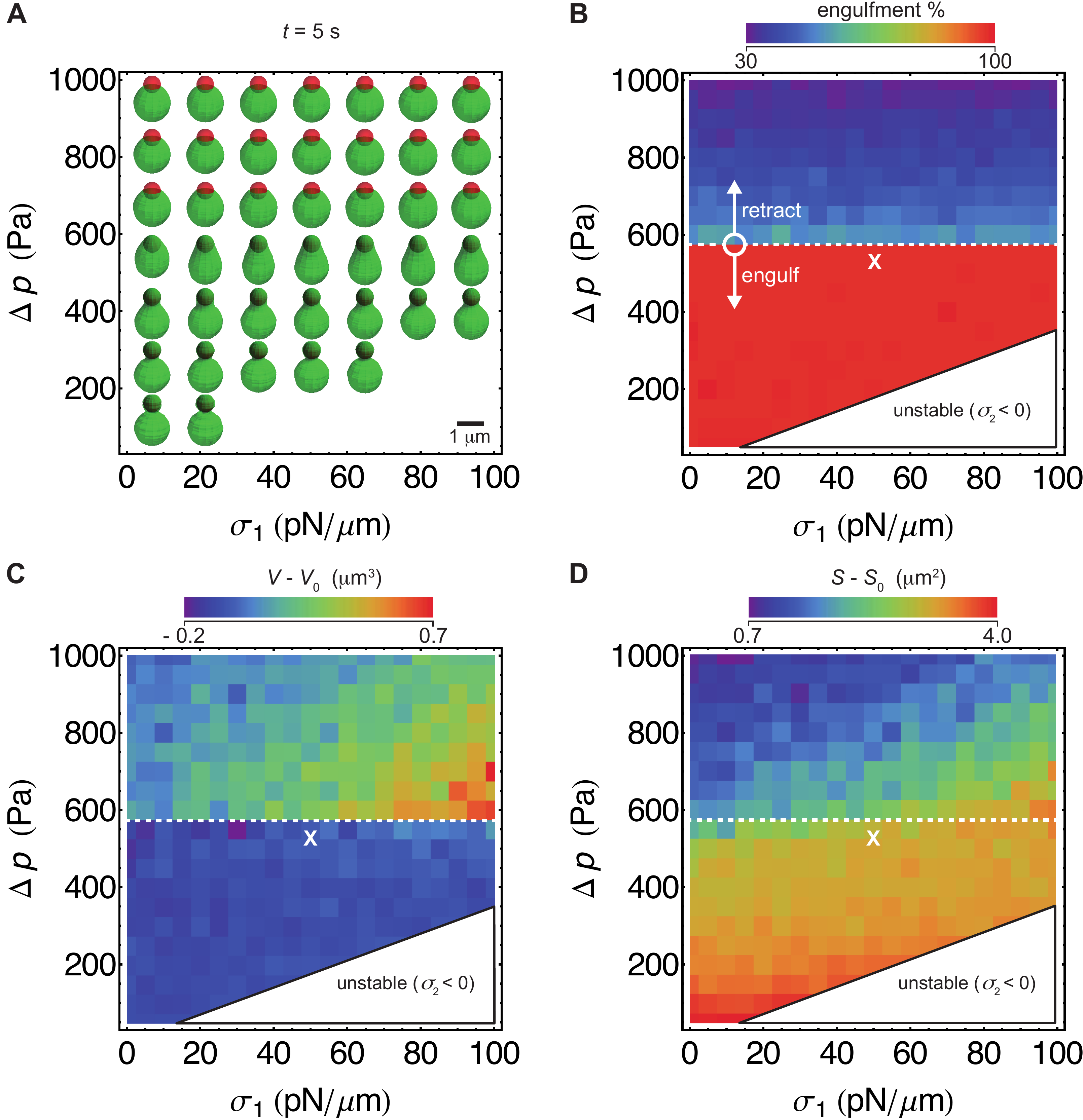}
\end{center}
\caption{
{\bf Simulation results for engulfment and volume/surface area changes.} (A-D) Simulation snapshots for different parameter combinations for surface tension ($\sigma_1$) and $\Delta p$ (pressure difference) at 5 s for fixed SpoIIQ-SpoIIIAH surface density $\rho = \rho_{\rm max}= 10^4\mu {\rm m}^{-2}$. Simulations that reached full engulfment earlier than 5 s were terminated and last snapshots are displayed. (B) Percentage of forespore-surface area enclosed by mother membrane. White dashed line separates regions of full and partial engulfment. White cross shows the parameters used for Fig. \ref{Fig_06}A and B. (C and D) Volume and surface area of mother cell.
}
\label{Fig_05}
\end{figure}

Using the model governed by Eq. \ref{Langevin1} we numerically simulated forespore engulfment in real time (Fig. \ref{Fig_05}, Movie S3). At the beginning of the simulations mother cell and forespore have spherical shape and intersect at a single point of their perimeters as shown in Fig. \ref{Fig_04}A. To better understand contributions of linear surface tension ($\sigma_1$) and pressure difference ($\Delta p$) on engulfment we varied them while keeping other parameters constant (for parameters see Text S1). We note that the explored surface tension ($\leqslant$ 100 pN/$\mu$m) is smaller than the experimentally observed rupture tension ($\sim$ 20 nN/$\mu$m)\cite{Needham1990,Evans:2003} and that explored pressure differences ($\leqslant$ 1000 Pa) are suitable for an osmotically balanced medium (see Materials and Methods). 

In Fig. \ref{Fig_05}A we show simulation snapshots at $5$ s. Simulations that reached full engulfment earlier than 5 s were terminated and last snapshots were displayed. In Fig. 5B the white dashed line separates successful engulfment ($\Delta p<$ 575 Pa) from retraction ($\Delta p>$ 575 Pa). In the region of successful engulfment with $\Delta p<$ 200 Pa and $\sigma_1<$ 30 pN/$\mu$m we observed engulfment with thin, tight cups as experimentally observed (Fig. \ref{Fig_05}A and B). In the region between retraction and thin cups we also observed successful engulfment but with adiabatic broad cups resulting in almost spherical mother cells.

In the region of thin cups engulfment is fast, taking only $\sim$ 1-2 s which is even faster than $\sim$ 1-2 min from experiments  (Fig. \ref{Fig_02}). We attributed this time discrepancy to the limiting factor of cell-wall removal in the experimental setup, since residues of the cell wall can prevent significant membrane fluctuations in our simulations, therefore delaying engulfment (Fig. S4). Indeed, movies of engulfment show elongated cells even minutes after addition of lysozyme \cite{Broder:2006}. For example, in Fig. \ref{Fig_01}A, top, at  $0$ min (the onset of volume loss) the engulfing mother cell still has elongated shape even though this is $6$ min after lysozyme treatment. To explore possible simulated engulfment times we varied the kinematic parameter $\eta_{\rm med}$ that represents the effective medium viscosity. This parameter sets the time scale of engulfment but does not influence the morphology of the engulfing cup (Fig. S5). We found that even for extremely high, experimentally observed $\eta_{\rm med}$, simulated engulfment times are still about an order of magnitude smaller than the experimentally observed engulfment times (Fig. S5).

To better understand physical changes of the simulated mother cell during engulfment, we plotted volume and surface area relative to their initial values (Fig. \ref{Fig_05}C and D). In the parameter region of successful engulfment we observed a mother-cell volume loss of $\sim$ 0.2 $\mu$m$^3$, which corresponds to the forespore volume. However, the experimentally observed mother-cell volume loss (Fig. \ref{Fig_02}A) is about $\sim$ 2 times higher. Furthermore, simulated surface areas increase in this region, while experimentally observed surface areas decrease slightly (Fig. \ref{Fig_02}B). This discrepancy may be explained by effects not included in our simulations, e.g. cytosol leakage (see Discussion).

\subsection*{Bistable engulfment is sensitive to zipper-protein surface density}

\begin{figure}[!ht]
\begin{center}
\includegraphics[width=12cm, trim = 0cm 0cm 0cm 0cm, clip=true]{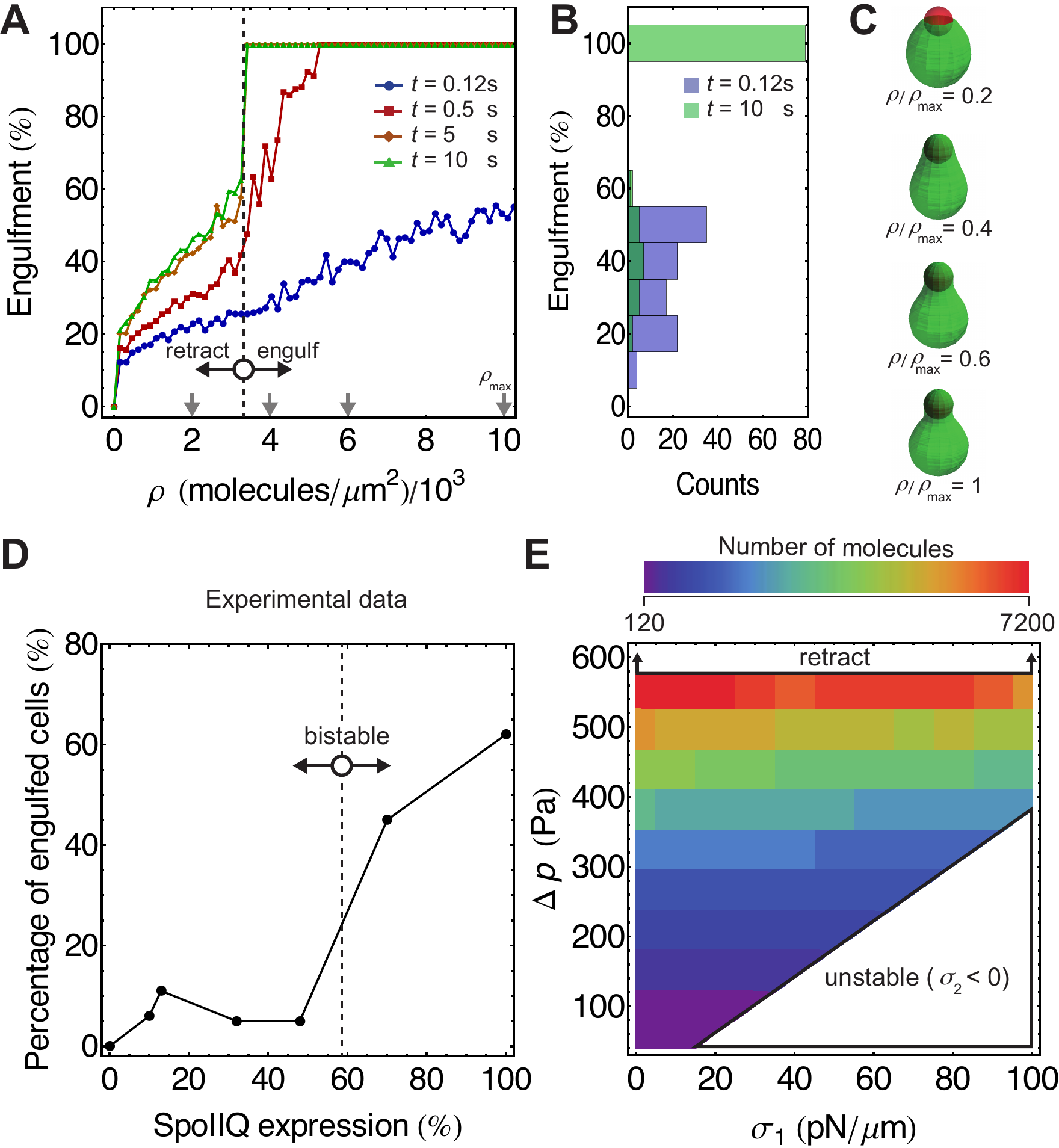}
\end{center}
\caption{
{\bf Bistable engulfment depends on zipper-molecule density.} (A) Engulfment as a function of SpoIIQ protein surface density for $\sigma_1$ = 50 pN/$\mu$m and $\Delta p$ = 500 Pa. The total binding energy was converted to SpoIIQ protein-surface density using the binding energy of a single SpoIIQ-SpoIIIAH bond (see Text S1) \cite{Meisner:2011}. Consistent with experimental results shown in (D) (extracted from \cite{Broder:2006}), engulfment is  threshold-dependent on number of SpoIIQ proteins expressed in forespores. Gray vertical arrows point to surface densities for which snapshots are shown in (C) at 5 s. (B) Simulations lead to bistable outcome at later times ($t$ = 10 s) with two subpopulations of stalled and fully completed cups. (E) For each set of constraint parameters ($\sigma_1$ and $\Delta p$) we performed a surface-density scan as in (A). The lower bound on the critical number of SpoIIQ molecules necessary for engulfment ranges from $\sim$ 120 to $\sim$ 7200 molecules depending on constraint parameters. 
}
\label{Fig_06}
\end{figure}

Experiments showed that engulfment depends critically on the number of expressed SpoIIQ zipper protein in a threshold-like fashion (Fig. \ref{Fig_06}) \cite{Broder:2006}; for SpoIIQ expression levels below a critical value mother-cell membrane retracts while for expression levels above the critical value the forespore is successfully engulfed (Fig. 6C). For wild-type cells with naturally occurring expression-level variation this presumably leads to the observed 60/40 $\%$ bistable engulfment outcome \cite{Broder:2006}. 

To numerically determine the number of SpoIIQ proteins required for successful engulfment, we varied the protein-surface concentration ($\rho$) for model parameters $\sigma_1$ = 50 pN/$\mu$m and $\Delta p$ = 500 Pa (Fig. \ref{Fig_06}A and B). Since in experiments the expression of SpoIIQ was reduced compared to wild-type cells, we assumed that the pool of SpoIIIAH molecules is large compared to SpoIIQ. Therefore, the limiting factor in engulfment is attributed to the surface concentration of SpoIIQ proteins. In Fig. \ref{Fig_06}A successful engulfment occurs for SpoIIQ surface densities above 3350 $\mu$m$^{-2}$. Since forespore surface area is $\sim$ 2 $\mu$m$^2$ (Fig. \ref{Fig_02}B), this produces a lower bound of $\sim$ 6700 SpoIIQ molecules in the forespore necessary for engulfment. To further explore the role of physical constraints on the number of critical zipper proteins that lead to successful engulfment, we scanned the parameter space for successful engulfment in Fig. \ref{Fig_05}B and found lower bounds on the number of molecules indispensable for engulfment (Fig. \ref{Fig_06}D). This lower bound can be as low as $\approx$ 120 SpoIIQ molecules for $\sigma_1<$ 30 pN/$\mu$m and $\Delta p<$ 200 Pa and increases to $\sim$ 7200 SpoIIQ molecules for $\Delta p>$ 550 Pa. Therefore, our simulations predict that under certain osmotic conditions and surface tensions, engulfment can occur with only $\sim$ 100 SpoIIQ molecules.

\subsection*{Experimental test of model predictions}
\begin{figure}[!ht]
\begin{center}
\includegraphics[width=15cm, trim = 0cm 0cm 0cm 0cm, clip=true]{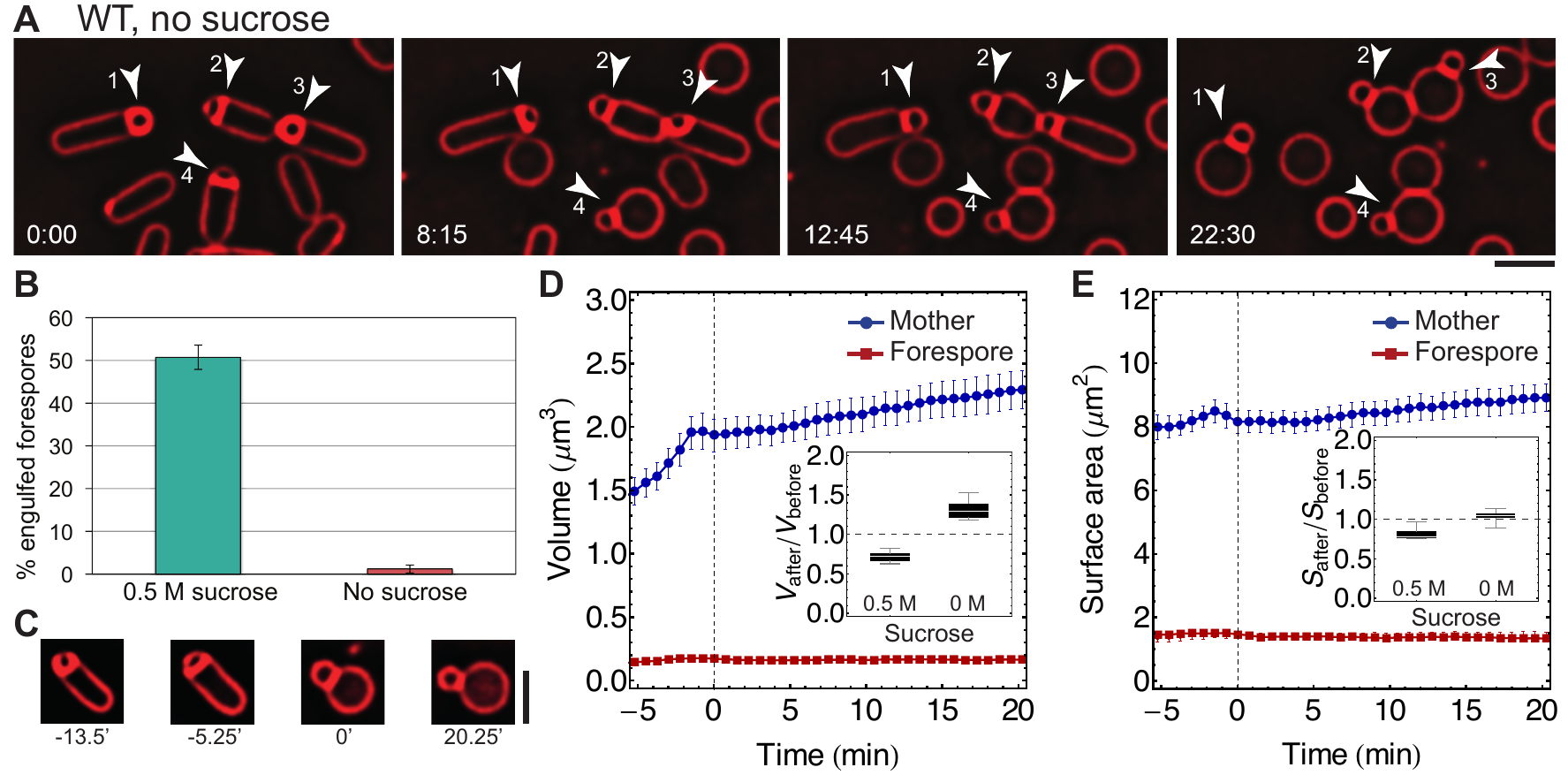}
\end{center}
\caption{
{\bf Decreasing medium osmolarity causes retraction as theoretically predicted.} (A) Medial focal plane images of sporulating cells treated with lysozyme in osmotically protected medium but without sucrose (see Materials and Methods, Movie S4). Fully engulfed cells retract (cells 1 and 3) while partially engulfed cells (cells 2 and 4) fail to engulf and undergo retraction as well. (B) Percentage of forespore engulfment after lysozyme treatment in medium with 0.5 M sucrose and without sucrose. Average +/- STD of four different microscopy fields,  each containing 136-160 cells in sucrose condition and 93-173 cells without sucrose. (C) Images of typical cell undergoing swelling and retraction. Time 0 minutes corresponds to fully rounded mother cell. (D) Mother-cell and forespore volume aligned in time based on 0 time point defined as in (C). After mother cells round, volume steadily increases. (Inset) Volume comparison for experiments with and without sucrose ($p < 10^{-8}$). $V_{\rm after}$ is mother-cell volume $\sim$ 2 minutes after engulfment or retraction corresponding to 0.5 M surose or no sucrose, respectively. $V_{\rm before}$ is initial mother-cell volume $\sim$ 8 minutes before transition. (E) Mother-cell and forespore surface area. (Inset) Surface area comparison for cases with and without sucrose ($p < 10^{-5}$).  Average +/- SEM of $n$ = 12 cells in panel (D, E). Scale bars 2 $\mu$m.
}
\label{Fig_new}
\end{figure}

Our model predicts that high osmotic pressure differences across the mother-cell membrane lead to swelling, reduced membrane fluctuations and membrane retraction, therefore preventing the completion of engulfment (Fig. \ref{Fig_05}A and B). To assess the validity of these predictions, we followed the progression of engulfment in absence of the cell wall for different osmotic conditions by time-lapse fluorescence microscopy (see Materials and Methods). Briefly, we resuspended sporulating cells in SMM buffer, either with 0.5 M sucrose to support protoplast engulfment (Fig. 1A, Movie S1)\cite{Broder:2006} or without sucrose, leading to an increase in the osmotic pressure difference between the cytoplasm and the extracellular medium. We then stained cell membranes with FM 4-64, added lysozyme to remove the cell wall, and performed time-lapse microscopy as the cell wall was degraded (Fig. \ref{Fig_new}A). In the presence of 0.5 M sucrose, more than 50 $\%$ of the cells completed engulfment as previously reported (Fig. \ref{Fig_new}B, Movie S1; \cite{Broder:2006}). However, in absence of sucrose the engulfing membrane retracted for almost all cells, and less than 2 $\%$ of them completed engulfment (Fig. \ref{Fig_new}A and B, Movie S4). Furthermore, membrane retraction was accompanied by an increase in mother-cell volume and surface area of $\sim$ 30 $\%$ and $\sim$ 5 $\%$, respectively (Fig. \ref{Fig_new}D and E). Hence, our data confirm the predictions of our biophysical model on the response to changes in medium osmolarity.

\section*{Discussion}

In this work we presented image analysis and modeling of forespore engulfment in the absence of the cell wall. Image analysis showed that engulfment occurs extremely fast ($\sim$ 1-2 min compared to $\sim$ 45 min with the cell wall), accompanied by a drastic volume loss ($\sim$ 35 $\%$) of the mother cell. During engulfment the initial broad cup dynamically changes to a thin cup, forming high-curvature membrane folds at the intersection between mother cell and forespore. Using Langevin simulations we showed that a Brownian ratchet model reproduces fast, out-of-equilibrium engulfment. Additionally, we numerically determined regions of engulfment and retraction, and predicted the number of SpoIIQ molecules necessary for successful engulfment. Similar, out-of-equilibrium Brownian ratchet mechanisms were previously used to explain molecular-motor directional movement \cite{Astumian:1998}, Lysteria motility from actin comets  \cite{Mogilner:1996,Zhu:2012}, filapodia protrusion \cite{Peskin:1993}, and unidirectional movement of other microscopic objects \cite{Di:2010}. 

Our model makes a number of predictions. The phase diagram in Fig. \ref{Fig_05}B predicts engulfment success and mother-cell morphology for a wide range of surface tension s($\sigma_1$) and pressure differences ($\Delta p$). For example, for a given  SpoIIQ surface density, high surface tension restricts the engulfment region while high osmotic pressure prevents engulfment. To test this prediction we increased the osmotic pressure difference by lowering the osmolarity of the suspended buffer. Decreased osmolarity indeed caused mother-cell swelling and stopped engulfment in line with our predictions (Fig. 7). Another prediction, such as the need for an excess of mother-cell membrane, might be tested by controlling the production of FapR, the major lipid homeostasis regulator \cite{Schujman:2003,Zhang:2008}.

There are a number of model limitations, which raise interesting issues. Based on our current model there is no stabilizing force for maintaining high curvatures (thin cups) once engulfment is accomplished. Therefore, the restoring constrain forces ($\textbf{F}^{\sigma}_{i}+\textbf{F}^{\Delta p}_{i}$) and bending force ($\textbf{F}^{\mathrm {bend}}$) will flatten high curvatures within $\sim$ 1 s on the length scale of 1 $\mu$m (see Fig. S3C and Eq. \ref{taun}). To estimate the force necessary to prevent membrane flattening we measured restoring forces in the neck region after engulfment is completed (Fig. S6). Typical radial forces were about 10 pN. Three factors can contribute to experimentally observed ``snowman-like'' shapes at late stages of engulfment (Fig. \ref{Fig_01}A, top): residues of the mother cell wall, potential outer/inner membrane binding, and lipid/protein sorting. These factors could be investigated experimentally. First, fluorescent cell-wall labeling \cite{Kuru:2012,Liechti:2013} could rule out possible remnants of cell wall after lysozyme treatment, which may preserve high membrane curvatures (see also Fig.  S4). Second, thin cups could also form if an unknown cohesive factor binds to outer and inner cup membrane thus preventing their separation. Third, membrane lipids that localize to high negative curvatures \cite{Kawai:2004,Renner:2011,Govindarajan:2013} together with curvature sensing proteins such as SpoVM \cite{Ramamurthi:2009} could contribute to the formation of stable structures preventing membrane flattening. To better explore this stabilizing mechanism we simulated engulfment with membranes that have positive and negative intrinsic curvatures $c_0$ (Fig. S7). We found that successful engulfment proceeds in range (-120 $\leqslant c_0 \leqslant$ 20) $\mu$m$^{-1}$. Interestingly, high negative curvatures prevent membrane flattening by forming tight and thin cups in the neck region typical for the snowman shape (Fig. S7, bottom, left). Therefore, the stability of tight cups after engulfment completion should be a topic of future experimental investigation, leading to better insights and advancement in modeling. 

Our current model fell short in reproducing the exact volume loss and surface-area conservation. To numerically find parameter regimes that lead to thin cups and observed volume and surface-area changes, we constructed a new albeit less physically grounded model with explicit volume and surface-area constraints (see Fig. S8 and Text S1). Using this model we determined the parameter region in which experimentally observed volume and surface-area changes occur similar to Fig. \ref{Fig_05}. As a result, engulfment occurs when either volume or surface area is not conserved as previously proposed \cite{Lizunov:2006} (see Movie S3). Volume loss may contribute to engulfment by effectively decreasing the surface tension, therefore boosting excess membrane for forespore engulfment. However, by close inspection of sporulating cells, we observed leakage of the cytosol in one of the cells after membrane removal at the onset of volume loss (Fig. S9). This direct loss of cytosol was not explicitly included in our models. One plausible interpretation of volume loss can be attributed to the hypertonic solution of suspended buffer containing 0.5 M of sucrose and producing an osmotic pressure of $\sim$ 12 atm \cite{Broder:2006}. Although this pressure is comparable to osmotic pressures inside of bacteria to balance pressures across the membrane, water leakage from bacteria and/or partial lyses after cell-wall removal is expected \cite{Koch1984,Misra:2013}. Therefore, future experiments may help address this issue.

An important future goal is the theoretical understanding of forespore engulfment in the presence of the cell wall, including membrane fission as the last stage of engulfment \cite{Doan:2013}. High turgor pressure and constraints from the cell wall and septum must provide difficult constraints for the engulfing mother cell, partially explaining the long engulfment time. In the presence of the cell wall, it has been proposed that peptidoglycan hydrolysis and new cell-wall deposition play major roles in the leading-edge membrane movement around the forespore \cite{Abanes:2002,Blaylock:2004,Doan:2005,Morlot:2010,Meyer:2010,Rodrigues:2013}. First, membrane proteins SpoIID, SpoIIM, and SpoIIP (DMP) form a complex and localize to the leading edge of the moving membrane. Since SpoIID and SpoIIP degrade peptidoglycans \cite{Abanes:2002,Chastanet:2007} and play an important role for thinning the septum, it has been proposed that the DMP complex is a processive motor for membrane advancement. However, a mechanistic description of this motor is still missing. Second, new cell-wall deposition at the leading edge may provide an additional motor-like mechanism for membrane movement \cite{Meyer:2010,Tocheva:2013}. A similar mechanism was proposed for cytokinesis of fission yeast \textit{Schizosaccharomyces pombe}, where polymerizing septum fibrils contribute to inward septum ingression. Using Brownian ratchet modeling of this process it was estimated that a single $\beta$-glucan fibril can exert polymerization force of $\sim$ 10 pN  \cite{Proctor:2012}. Together, future modeling will produce a better understanding of the complicated process of forespore engulfment and decipher the contributions from each mechanism towards the total force at the leading membrane edge.

In conclusion, our quantitative model of engulfment in absence of the cell wall provided first insights into mother-cell morphologies, such as cup shape, engulfment dynamics, and bistability. Due to simplicity from cell-wall removal, energy limitations, and absence of cytoskeletal cortex, forespore engulfment could present a minimal system for studying bistability and decision-making. Interestingly, bistability and commitment to engulf were previously identified in phagocytosis \cite{Zhang:2010,Tollis:2010}, thus showing similarities to forespore engulfment. This similarity becomes enhanced, if we speculate that the SpoIIQ-SpoIIIAH backup mechanism may instead be the original core mechanism, which may have evolved before the more complex DMP-based mechanism. In fact, the C-terminal domain of SpoIIIAH is homologous to YscJ/FliF protein family forming multimeric rings in type-III secretion system and flagella motors \cite{Meisner:2008}, pointing towards an ancient mechanism. General biophysical principles may also apply to other types of engulfment including the penetration of the red blood cell by the malaria parasite \cite{Cowman:2006}.

\section*{Materials and Methods}

\subsection*{Image analysis}
We used the semi-automated active contour software JFilament \cite{Smith:2010} available as ImageJ plugin to extract the membrane position over time. All movies analyzed were previously published in \cite{Broder:2006}. The information about membrane positions obtained from medial focal plane is used to obtain 3D volume and surface area by assuming rotational symmetry around the axis connecting center of mass of mother cell and forespore. Kymographs as in Fig. \ref{Fig_02}C were created by collecting intensities along the forespore contours using JFilament. Subsequently, pixel angles were determined using pixel position relative to the mother-forespore frame as defined in inset of Fig. \ref{Fig_02}C. To test the image analysis method, which was used to estimate the number of membrane folds in the neck region of cup (Fig. \ref{Fig_03}), we measured fluorescence intensities along known single ($I_{\rm single}$) and double ($I_{\rm double}$) membranes (Fig. S10). This analysis produced $I_{\rm double}/I_{\rm single}$ = 2.3 $\pm$ 0.6 as expected.

\subsection*{Cross-correlation analysis}
To compare two signals $a(t_i)$ and $b(t_i)$ that are given at discrete time points $t_i$, we calculated the cross-correlation function: 
\begin{equation}
 C_{\rm cross}(\Delta t) =\frac{\sum_{i=0}^{N} (a(t_i) - \bar{a})(b(t_i - \Delta t)-\bar{b})}{\sqrt{ \sum_{i=0}^{N} (a(t_i) - \bar{a})^2 }  \sqrt{ \sum_{i=0}^{N} (b(t_i) - \bar{b})^2 }}.
\label{cross}
\end{equation}
Here, $\bar{a}$ and $\bar{b}$ are the average signals, and $N$ is the total number of discrete time points. 

\subsection*{Langevin simulation}
A stochastic Langevin equation is used to simulate mother-cell membrane dynamics (see Eq. \ref{Langevin1}). Simulation time step and distance between neighboring beads were ${\rm d}t = 0.5$ $ \mu$s and $l_{0}$ = 10 nm, respectively. When the distance between two neighboring beads exceeded the equilibrium distance by $\pm$25 $\%$ the whole contour was rebeaded using a linear-interpolation method \cite{Smith:2010}. Additionally, membrane-excluding volume is also implemented; whenever the distance between two beads that are not nearest neighbors was less than $l_{0}$ a repelling radial force of 50 pN was applied to both beads. Analytical details of the model of Eq. \ref{Langevin1} are explained in the following sections.


\setlength{\parskip}{10pt plus 1pt minus 1pt}
\noindent
\textbf{Membrane bending}: The bending energy corresponding to the surface area ($\Delta A_i$) of the $i^{\rm th}$ ribbon is given by \cite{Boal:2002}:
\begin{equation}
E^{\mathrm {bend}}_{i} = \frac{1}{2} \kappa_{\mathrm {b}} (c_{\mathrm m,i} + c_{\mathrm p,i})^2   \, \Delta A_i,
\label{Ebend}
\end{equation}
where $c_{\mathrm m,i} \equiv\, {\mathrm d \theta_i}/{\mathrm d s}$ is the meridian principle curvature, $c_{\mathrm p,i} \equiv \, {\sin \theta_i }/{x_i}$ is the principle curvature along the parallels \cite{Deuling:1976}, $l_{0}$ is the distance between two neighboring beads, $\kappa_{\mathrm {b}}$ is the membrane bending rigidity, and  $\theta_i$ is the angle between unit normal vector ${\hat n}_i$ of the contour and $z$-axis (Fig. \ref{Fig_04}A). Here, we neglected the Gaussian curvature contributions as the topology of the mother-cell membrane does not change during engulfment \cite{Boal:2002}. The role of intrinsic membrane curvatures on cup shapes is explored in Fig. S7. 
Summing Eq. \ref{Ebend} over the whole surface area, we obtain the total bending energy:
\begin{equation}
E^{\mathrm {bend}}_{\mathrm {tot}} =\pi \kappa_{\mathrm {b}} l_0 \sum_{j=1}^{N} x_j (c_{\mathrm m,j} + c_{\mathrm p,j})^2,
\label{Ebend1}
\end{equation}
where $N$ is the total number of beads. Therefore, the force due to bending energy is given by:
\begin{equation}
\textbf F^{\mathrm {bend}}_{i} =-\frac{\partial E^{\mathrm {bend}}_{\mathrm {tot}}}{\partial \textbf{R}_i}.
\label{Fbend}
\end{equation}
For detailed analytical derivation of each component of the bending force see Text S1. 

\setlength{\parskip}{10pt plus 1pt minus 1pt}
\noindent
\textbf{Stochastic force}: The stochastic force due to thermal noise is defined as \cite{Pasquali:2002}:
\begin{equation}
\langle \textbf F^{\mathrm {stoch}}_{i} \textbf F^{\mathrm {stoch} \, \rm T}_{i} \rangle = \frac{2k_{\rm B}T \zeta_{0} (\frac{L_i}{l_0})^2}{\Delta t} \hat {\textbf I},
\label{fstoch}
\end{equation}
where $k_{\rm B}T$ is the thermal energy, $\Delta t$ the simulation time step, $\zeta_{0}$ is the frictional coefficient of the segment with length $l_{0}$, and $\hat {\textbf I}$ is the unit matrix. Here, we introduced the $({L_i}/{l_0})^2$ correction, making displacements due to the thermal noise independent of position $x_i$ and $z_i$. As a result, all beads in thermal equilibrium fluctuate with average zero displacement under a force whose variance is given by Eq. \ref{fstoch}. In Fig. S2 we numerically determined the effective temperature so that the typical amplitude of fluctuations is of the order $\sim$ 15 nm as observed \cite{Sackmann:1996,Betz:2009}. 

\setlength{\parskip}{10pt plus 1pt minus 1pt}
\noindent
\textbf{Mother-forespore zipper binding}: SpoIIQ and SpoIIIAH protein binding is modeled with a spring-like interaction with interaction range $l_{\rm {QAH}}$: 
\begin{equation}
\textbf F^{\mathrm {QAH}}_{i} = - k_{\mathrm {QAH}} \cdot (\lvert \textbf R_{i} - \textbf R_{\mathrm {fspore}} \rvert - l_{\mathrm {QAH}}) \, \cdot \mathrm H ( l_{\rm {QAH}} - \lvert \textbf R_{i} - \textbf R_{\mathrm {fspore}} \rvert)
\label{spoiiq_spoiiiah}
\end{equation}
Here, $\textbf R_{\mathrm {fspore}}$ is the position of forespore contour closest to the mother's $i^{\rm th}$ bead, $l_{\rm {QAH}}$ is the interaction distance between SpoIIQ-SpoIIIAH, $k_{\rm {QAH}}=2 E_{\rm {QAH}}/l_{\rm {QAH}}^2$ is the spring constant of SpoIIQ-SpoIIIAH interaction, and $\rm H( \, )$ is the Heaviside step function preventing binding distances larger than $l_{\rm {QAH}}$.

\setlength{\parskip}{10pt plus 1pt minus 1pt}
\noindent
\textbf{Surface tension and pressure difference}:
Energy terms from surface tension ($\sigma_1$ and $\sigma_2$) and osmotic pressure difference ($\Delta p$) balance each other in thermal equilibrium \cite{Derenyi:2002,Deuling:1976}.
\begin{equation}
\tilde{E} \equiv E^{\sigma} + E^{\Delta p} = \sigma_1 S (1+\sigma_2 S) -\Delta p V ,
\label{Etildi}
\end{equation}
where $S$ is the mother-surface area and $V$ is the mother-cell volume (see Fig. S11). The non-linear term $\sigma_1 \sigma_2 S^2$ is used to stabilize the initial mother cell that was assumed equilibrated at the onset of engulfment \cite{Herant:2006}. This non-linear term allowed independent exploration of parameters $\sigma_1$ and $\Delta p$ while parameter $\sigma_2$ is derived from the equilibrium condition $\frac{\partial {\tilde{E}}}{\partial S}\vert_{S=S_0}= 0$, where $S_0$ is the initial mother-cell area (for detailed analytical derivation see Text S1). Finally, the forces are obtained as
in Eq. \ref{Fbend}:
\begin{equation}
\textbf{F}^{\sigma}_{i}+\textbf{F}^{\Delta p}_{i} =-\frac{\partial \tilde{E}}{\partial \textbf{R}_i}.
\label{Ftilda}
\end{equation}

\subsection*{Model validation}
To validate our biophysical membrane model, we quantified thermal fluctuations from simulations and compared this with analytical results from \cite{Milner:1987,Meleard:1992,Pecreaux:2004} (see Fig. S3). Fourier mode with wave number $n$ is calculated as in \cite{Pecreaux:2004}:
\begin{equation}
\tilde{c}_n = \frac{1}{\pi \langle R \rangle ^2}  \int_{-\pi R}^{\pi R} (R(s)-\langle R \rangle) \mathrm{e}^{-i n \frac{s}{\langle R \rangle}}\,\mathrm{d}s,
\label{Fourier}
\end{equation}
with $\langle R \rangle$ the average membrane radius.

\setlength{\parskip}{10pt plus 1pt minus 1pt}
\noindent
Fourier modes are collected during first 10 s of simulations. Autocorrrelation function ($C_{c_n}$) of each Fourier mode is calculated and fitted to exponential function ($C_{c_{n}} \sim {\rm e}^{-t/\tau_{n}}$) \cite{Duwe:1990}, where $\tau_{n}$ is the relaxation time for each Fourier mode that satisfies analytical expression \cite{Milner:1987,Meleard:1992,Betz:2012}:
\begin{equation}
\tau_n = \frac{\eta_{\rm med} \langle R \rangle ^3}{\kappa_{\rm b}} \frac{2n+1}{(n+2)(n-1) ( \tilde{\sigma} +n(n+1) )} \frac{2n^2+2n-1}{n(n+1)},
\label{taun}
\end{equation}
with  $\tilde{\sigma} \equiv \frac{\sigma_1 \langle R \rangle ^2}{\kappa_{\rm b}}$ the reduced membrane tension, $\eta_{\rm med}$ the medium viscosity, and $\kappa_{\rm b}$ the bending stiffness.

\setlength{\parskip}{10pt plus 1pt minus 1pt}
\noindent
To validate membrane shapes in thermal equilibrium, we collected 6000 simulated membrane contours and calculated the variance for each Fourier mode, also known as \textit{dimensionless spectrum}. The analytical expression for dimensionless spectrum of planer membranes is given by \cite{Pecreaux:2004}:
\begin{equation}
S_{\rm pl}(n) \equiv \langle c_{n}^2\rangle-\langle c_{n} \rangle ^2 = \frac{1}{\pi \langle R \rangle ^3} \frac{k_{\rm B}T}{\sigma_1} \left( \frac{\langle R \rangle}{n}-\frac{1}{\sqrt{\frac{\sigma_1}{\kappa_{\rm b}} + \frac{n^2}{\langle R \rangle ^2}}} \right).
\label{spl}
\end{equation}

\subsection*{Time-lapse fluorescence microscopy}
\noindent
\textit{Bacillus subtilis} PY79 sporulation was induced by resuspension  at 37 $^{\circ}$ C similar to \cite{Sterlini:1969}, except that the bacteria were grown in 25 $\%$ LB prior to resuspension rather than in CH medium. Samples were taken two hours and forty-five minutes after resuspension, spun at 9000 rpm for 10 s, and resuspended either in 25 $\mu$l of SMM buffer (0.5 M sucrose, 20 mM maleic acid, 20 mM MgCl$_2$, pH 6.5) or in the same buffer without sucrose. 10 $\mu$l of the resuspended culture were placed on a poly-L-lysine-treated coverslip and mixed with lysozyme and FM 4-64 (final concentrations 1 mg/ml and 5 $\mu$g/ml, respectively). Pictures were taken at room temperature, every 45 seconds for one hour, using an Applied Precision optical sectioning microscope equipped with a Photometrics CoolsnapHQ$^2$ camera. Images were deconvolved and analyzed with SoftWoRx version 5.5 (Applied Precision) and ImageJ.

\section*{Acknowledgments}
We thank Morgan Beeby and Erdem Karatekin for critical reading of the manuscript.

\vspace{4cm}

\bibliography{sporulation}

\section*{Supporting information}
\textbf{Text S1.} Supporting information includes image analysis of cells lacking zipper proteins, details of numerical simulations, simulation parameters, and supplementary figures.

\noindent
\textbf{Movie S1. Engulfment with 0.5 M of sucrose.}  Medial focal plane of sporulating \textit{B. subtilis} cells treated with lysozyme two hours and forty-five minutes after resuspension. Membranes were stained withFM 4-64. Lysozyme was added just before imaging. Engulfing or retracting membranes are observed. Movie length 1h. (AVI)

\noindent
\textbf{Movie S2. Membrane folds along contours.} FM 4-64 channel (top left) and SpoIIIJ-GFP channel (top right) from time lapse fluorescence microscopy from \cite{Broder:2006}.  For a given cell and channel the average pixel intensity along the single mother membrane far from the boundary ($I_{\rm offb}$) is calculated for each time point (as in Fig. \ref{Fig_03}A). Intensities along all membranes are then scaled by this average intensity. Membrane folds along the contours are shown for  FM 4-64 channel (bottom left) and SpoIIIJ-GFP (bottom right) with color scale on the far right. Accumulation of membrane folds at the mother-forespore boundary accompanies engulfment process. Movie length 21.75 minutes. (AVI)

\noindent
\textbf{Movie S3. Langevin simulations.}  Simulations of Model 1 (left) and Model 2 (right). In the Model 1 engulfment occurs for $\Delta p<$ 575 Pa, while in Model 2 engulfment occurs when either volume or surface area is not conserved. Movie length 5 seconds. (AVI)

\noindent
\textbf{Movie S4. Membrane retraction without sucrose.} Protocol is similar as for Movie S1 but without addition of sucrose. Movie length 1h. (AVI)

\newpage

\section*{Figure Legends}

\vspace{5cm}
\vspace{2cm}


\end{document}